\documentclass[final,3p,12pt]{elsarticle}
\usepackage{amsfonts}
\usepackage{amssymb}
\usepackage{amsmath}
\usepackage{graphicx}

\def\sech{{\rm sech}}

\journal{Physics Letters A}

\begin{document}

\begin{frontmatter}

\title{Modulation instability and solitons in two-color nematic crystals}

\author{Theodoros P. Horikis}
\address{Department of Mathematics, University of Ioannina, Ioannina 45110,
Greece}

\begin{abstract}
The conditions under which stable evolution of two nonlinear interacting waves are derived within
the context of nematic crystals. Two cases are considered: plane waves and solitons. In the first
case, the modulation instability analysis reveals that while the nonlocal term suppresses the
growth rates, substantially, the coupled system exhibits significantly higher growth rates than
its scalar counterpart. In the soliton case, the necessary conditions are derived that lead the
solitons to exhibit stable, undistorted evolution, suppressing any breathing behavior and
radiation, leading to soliton mutual guiding.
\end{abstract}


\end{frontmatter}

\section{Introduction}

Nonlinear wave propagation in optical media is normally the result of a certain balance between
dispersion/diffraction and nonlinearity \cite{ablowitz}. This balance can lead to special structure
formations which are termed solitons or solitary waves. While these structures are proven to be
very stable, the necessary balance needed to form them is not always achieved. As a result, due to
the dominance of dispersive/diffractive spreading or nonlinear focusing an unwanted growth of
initial waves can be observed. This amplitude modulation is widely known and studied in the context
of nonlinear Schr\"odinger equation (NLS) systems \cite{agrawal1,drummond}.

The NLS system, however, for several physically relevant contexts turns out to
be an oversimplified description as it cannot model, for example, gain and loss
which are inevitable in any physical system \cite{agrawal,kivshar_book}. Hence,
in order to model important classes of physical systems in a relevant way, it
is necessary to go beyond the standard NLS description. To this end it is
important to consider adding higher-order terms into the NLS equation, so as to
accurately describe waves in dissipative nonlinear systems. In fact, for
specific classes of these higher-order terms terms the resulting system may
still be  integrable \cite{yang} and one can utilize the mathematical tools
also provided for the simple NLS case.

In some physical systems simply adding higher order and/or dissipative terms will not result in a
better description of their properties. These are, for example, systems with different nonlinear
response such as nonlocal. Nonlocality refers to the reorientational response of the medium to the
pulses having a width much greater than the width of the light beams. One of the most interesting
materials with nonlocal nonlinear response are nematic liquid crystals. In this context,
nonlocality is highly tuneable and solitons, termed nematicons, have been observed induced by a
regular local nonlinearity, a nonlocal nonlinearity or a combination of both \cite{skuse}.
Furthermore, nonlinear nonlocal equations similar to those governing nematicons have been found to
govern solitary waves in other media, for instance, colloidal \cite{matuszewski}, thermal
\cite{kroli2}, photorefractive \cite{segev} and plasmas \cite{litvak}.

One of the fundamental effects of wave propagation in nonlinear media is modulation instability
(MI). MI was first described in the context of water waves and is referred to as the Benjamin-Feir
instability \cite{bf}. This is the instability of nonlinear plane waves against small perturbations
resulting in their exponential growth. It has been observed experimentally and identified
mathematically in various physical applications (other than water waves) \cite{mi}.

Much like soliton (or coherent structure) formation, MI exists due to the interplay between
nonlinearity and dispersion/diffraction. Thus, it highly depends on the nature of the nonlinearity,
e.g. cubic, quartic, saturable, nonlocal, etc \cite{esbensen,kavitha}. Furthermore, the effect is
significantly enhanced in coupled systems \cite{kourakis}. We intend to examine the effects of MI
in nonlocal media and in particular in two-color nematicons. In these media (in the scalar case),
the nonlocal term seems to suppress the growth rate and the overall effect of MI
\cite{kroli,kroli2}. In addition, exact stationary (soliton) solutions of the coupled equation may
still exist under specific requirements. These are termed paired solitons since they specify the
intensity profiles of both beams and occur in pairs \cite{makha,fuente,djf1,kivshar_book}.

The article is divided in sections as follows: In Section 2 we define the
equations that govern propagation in these media and perform the MI analysis.
Interestingly, we find that the stability criteria are similar to those of the
relative NLS system and do not depend on the nonlocal parameter. We also
identify two critical values: the maximum growth rate and the critical
wavenumber which defines the range of wavenumber that can induce the
instability. By doing so, we also identify a critical value for the nonlocality
which corresponds to the minimum value needed to stabilize a continuous wave of
specific wavenumber. In Section 3, the soliton solutions are discussed and
appropriate conditions that relate the relative amplitudes, in order for
undistorted evolution to occur, are derived. Finally, we summarize our findings
in Section 4.

\section{Modulation instability analysis}

The governing equations that describe two polarised, coherent light beams of two different
wavelengths propagating through a cell filled with a nematic liquid crystal read, in
non-dimensional form \cite{vector,skuse2},
\begin{align}
  i\frac{{\partial u}}{{\partial z}} &+ \frac{{{d_1}}}{2}\frac{{{\partial ^2}u}}{{\partial {x^2}}} +
  2{g_1}\theta u = 0
  \label{nem1}\\
  i\frac{{\partial v}}{{\partial z}} &+ \frac{{{d_2}}}{2}\frac{{{\partial ^2}v}}{{\partial {x^2}}} +
  2{g_2}\theta v = 0
  \label{nem2}\\
  \nu \frac{{{\partial ^2}\theta }}{{\partial {x^2}}} &- 2q\theta  =  - 2({g_1}|u{|^2} +
  {g_2}|v{|^2})
  \label{nem3}
\end{align}
The variables $u$ and $v$ are the complex valued, slowly varying envelopes of the optical electric
fields and $\theta$ is the optically induced deviation of the director angle. Diffraction is
represented by $d_1, d_2$ and nonlinearity by $g_1,g_2$. Importantly, these variables are allowed
to vary their signs while all other constants are taken positive. When the signs of diffraction and
nonlinearity are opposite, i.e. $d_1g_1,d_2g_2<0$, the system is termed defocusing and focusing
otherwise. The location of the relative signs is not important; multiplying Eqs.
(\ref{nem1})--(\ref{nem2}) by $-1$ and changing $z\rightarrow -z$ moves the sign difference from
one place to the other. The nonlocality $\nu$ measures the strength of the response of the nematic
in space, with a highly nonlocal response corresponding to $\nu$ large. The parameter $q$ is
related to the square of the applied static field which pre-tilts the nematic dielectric
\cite{peccianti,alberucci,assanto3}. Note, that the above system corresponds to the nonlocal regime
with $\nu$ large, where the optically induced rotation $\theta$ is small \cite{assanto3}. We will
be using dimensionless parameters here because the system above may also be used to describe light
propagation in nonlocal media, in general. For nematic crystals, in particular, $d_1,g_1,d_2,g_2,q$
are $O(1)$ while $\nu$ is $O(10^2)$ \cite{skuse2,skuse3}.

In order to investigate the MI of a pair of coupled waves we first consider the continuous wave
(cw) solution of Eqs. (\ref{nem1})--(\ref{nem3}), i.e.
\[
u=u_0 e^{2i g_1 \theta_0 z},\quad v=v_0 e^{2i g_2 \theta_0 z},\quad \theta_0=\frac{g_1 u_0^2+g_2
v_0^2}{q}
\]
where $u_0$ and $v_0$ are real constants. Now consider a small perturbation to this cw solution
\begin{equation}
u(z,x)=[u_0 + u_1(z,x)]e^{2i g_1 \theta_0 z},\quad v(z,x)=[v_0 + v_1(z,x)]e^{2i g_2 \theta_0 z}
\label{eq0}
\end{equation}
which we insert into the system (\ref{nem1})--(\ref{nem3}). In order to simplify the analysis we
first solve Eqs. (\ref{nem1})--(\ref{nem2}) for $\theta$ and substitute in Eq. (\ref{nem3}). While
this eliminates one dependent variable it raises the overall order of the system and as such it
only proves useful in the MI analysis where plane wave solutions are investigated (as solutions of
a linear system). When solitons or exact solutions are the object of the analysis this is not
recommended. The linearized equations (where terms of order $u_1^2$ and $v_1^2$ or higher have been
neglected) for the small perturbing terms are found to be
\begin{align}
  4iq\frac{{\partial {u_1}}}{{\partial z}} - 2i\nu \frac{{{\partial ^3}{u_1}}}{{\partial
  z\partial {x^2}}}
  - {d_1}\nu \frac{{{\partial ^4}{u_1}}}{{\partial {x^4}}} + 2{d_1}q\frac{{{\partial
  ^2}{u_1}}}{{\partial {x^2}}}
  + 8g_1^2u_0^2({u_1} + u_1^*) + 8{g_1}{g_2}{u_0}{v_0}({v_1} + v_1^*) &= 0 \label{lin1}\\
  4iq\frac{{\partial {v_1}}}{{\partial z}} - 2i\nu \frac{{{\partial ^3}{v_1}}}{{\partial
  z\partial {x^2}}}
  - {d_2}\nu \frac{{{\partial ^4}{v_1}}}{{\partial {x^4}}} + 2{d_2}q\frac{{{\partial
  ^2}{v_1}}}{{\partial {x^2}}}
  + 8g_2^2v_0^2({v_1} + v_1^*) + 8{g_1}{g_2}{u_0}{v_0}({u_1} + u_1^*) &= 0 \label{lin2}
\end{align}
Notice that the terms involving $\nu$, that induce the nonlocality, are higher order derivatives;
without these terms the problem simply reduces to the linearized NLS problem. This means that their
contribution is expected to be highly nontrivial, as they produce higher order polynomials in the
dispersion relation; the effect of these terms will become more prominent below. Eqs.
(\ref{lin1})--(\ref{lin2}) admit solutions of the form
\begin{equation}
\hspace*{-1cm} u_1(z,x)=c_1 e^{i (k x-\omega  z)} + c_2 e^{-i (k x-\omega  z)},\quad
v_1(z,x)=c_3 e^{i (k x-\omega  z)} + c_4 e^{-i (k x-\omega  z)}
\label{plane}
\end{equation}
provided the dispersion relationship
\begin{equation}
p_1(k){\omega ^4} + p_2(k){\omega ^2} + p_3(k)=0
\label{disp}
\end{equation}
with
\begin{align*}
p_1(k) &= 16\left( {{k^2}\nu  + 2q} \right)\\
p_2(k) &= { - 4\nu \left( {d_1^2 + d_2^2} \right){k^6} - 8q\left( {d_1^2 + d_2^2} \right){k^4} +
64\left( {{d_1}g_1^2u_0^2 + {d_2}g_2^2v_0^2} \right){k^2}} \\
p_3(k) &= {d_1^2d_2^2\nu {k^{10}} + 2d_1^2d_2^2q{k^8} - 16{d_1}{d_2}\left( {{d_2}g_1^2u_0^2 +
{d_1}g_2^2v_0^2} \right){k^6}}
\end{align*}
Eq. (\ref{disp}) is a bi-quadratic and can be solved analytically to produce $\omega=\omega(k)$ as
\begin{equation}
\omega  =  \pm \sqrt {\frac{{ - {p_2}(k) \pm \sqrt {p_2^2(k) - 4{p_1}(k){p_3}(k)} }}{{2{p_1}(k)}}}
\label{omega}
\end{equation}
MI will be exhibited when this $\omega$ has complex solutions: their imaginary
part, Im$\{\omega\}$, will give the relative growth rate, as suggested by Eq.
(\ref{plane}). To classify the nature of $\omega$ (real or complex) we need to
solve a system of inequalities to ensure Eq. (\ref{disp}) only admits real
solutions, thus avoiding any exponential growth. In particular, there are three
polynomials in $k$ which one needs to prove are positive \cite{wiki}. This will
provide the appropriate conditions for stability. These are:
\begin{align*}
\Delta(k) &= 65536 k^{14} (2 q + \nu k^2)[ \nu^2(d_1^2 - d_2^2)^2 k^8+4q\nu (d_1^2 - d_2^2)^2
k^6\\
  &+ 4 (d_1^2 - d_2^2) (d_1^2 q^2 - 8 d_1 g_1^2\nu u_0^2 +
   d_2 (-d_2 q^2 + 8 g_2^2 v_0^2 \nu)k^4\\
   &- 64 q (d_1^2 - d_2^2) (d_1 g_1^2 u_0^2 - d_2 g_2^2 v_0^2)k^2
 + 256 (d_1 g_1^2 u_0^2 + d_2 g_2^2 v_0^2)^2]^2 Q_1(k)\\
P(k) &= 128 (2 q +\nu k^2) Q_2(k) \\
D(k) &= 64 (2 q +\nu k^2)[-4\nu (d_1^2 + d_2^2) k^6 -8q (d_1^2 + d_2^2) k^4
  + 64 (d_1 g_1^2 u_0^2 + d_2 g_2^2 v_0^2) k^2]^2 Q_3(k)
\end{align*}
Hence it is sufficient to show that the polynomials
\begin{align*}
  Q_1(k) &= (d_1^2d_2^2\nu) {k^4} + (2d_1^2d_2^2q){k^2}
  - 16{d_1}{d_2} ( {{d_2}g_1^2u_0^2 + {d_1}g_2^2v_0^2}) \\
  Q_2(k) &= \nu( {d_1^2 + d_2^2} ) {k^4} + 2 q( {d_1^2 + d_2^2} ){k^2}
  - 16 ( {{d_1}g_1^2u_0^2 + {d_2}g_2^2v_0^2} )\\
  Q_3(k) &= {\nu ^2}{(d_1^2 - d_2^2)^2}{k^8} + 4 q\nu{(d_1^2 - d_2^2)^2} {k^6}\\
  &+ 4(d_1^2 - d_2^2)[d_1^2{q^2} - 8{d_1}g_1^2u_0^2\nu  + {d_2}( - {d_2}{q^2} + 8\nu g_2^2v_0^2)]{k^4}\\
  &- 64(d{1^2} - d{2^2})q({d_1}g_1^2u_0^2 - {d_2}g_2^2v_0^2){k^2}
  + 256{({d_1}g_1^2u_0^2 - {d_2}g_2^2v_0^2)^2}
\end{align*}
are always positive. This is obtained through Sturm's theorem \cite{sturm}; since the coefficients
of the highest order terms are positive and of even degree, it is sufficient to demand that these
polynomials do not exhibit real roots. This happens when:
\begin{equation}
(\mathrm{i})\; \frac{{g_1^2u_0^2}}{{{d_1}}} + \frac{{g_2^2v_0^2}}{{{d_2}}} < 0
\quad \mathrm{and} \quad
(\mathrm{ii})\; d_1g_1^2u_0^2+d_2g_2^2v_0^2<0
\end{equation}
It is now trivial to show that for both conditions to hold it is sufficient to pose that both $d_1$
and $d_2$ are negative. As such, the coupled system also follows the condition of stability of the
single equation and stability is achieved iff the system is fully defocusing.

\begin{figure}[ht]
\centering
\includegraphics[height=5.5cm]{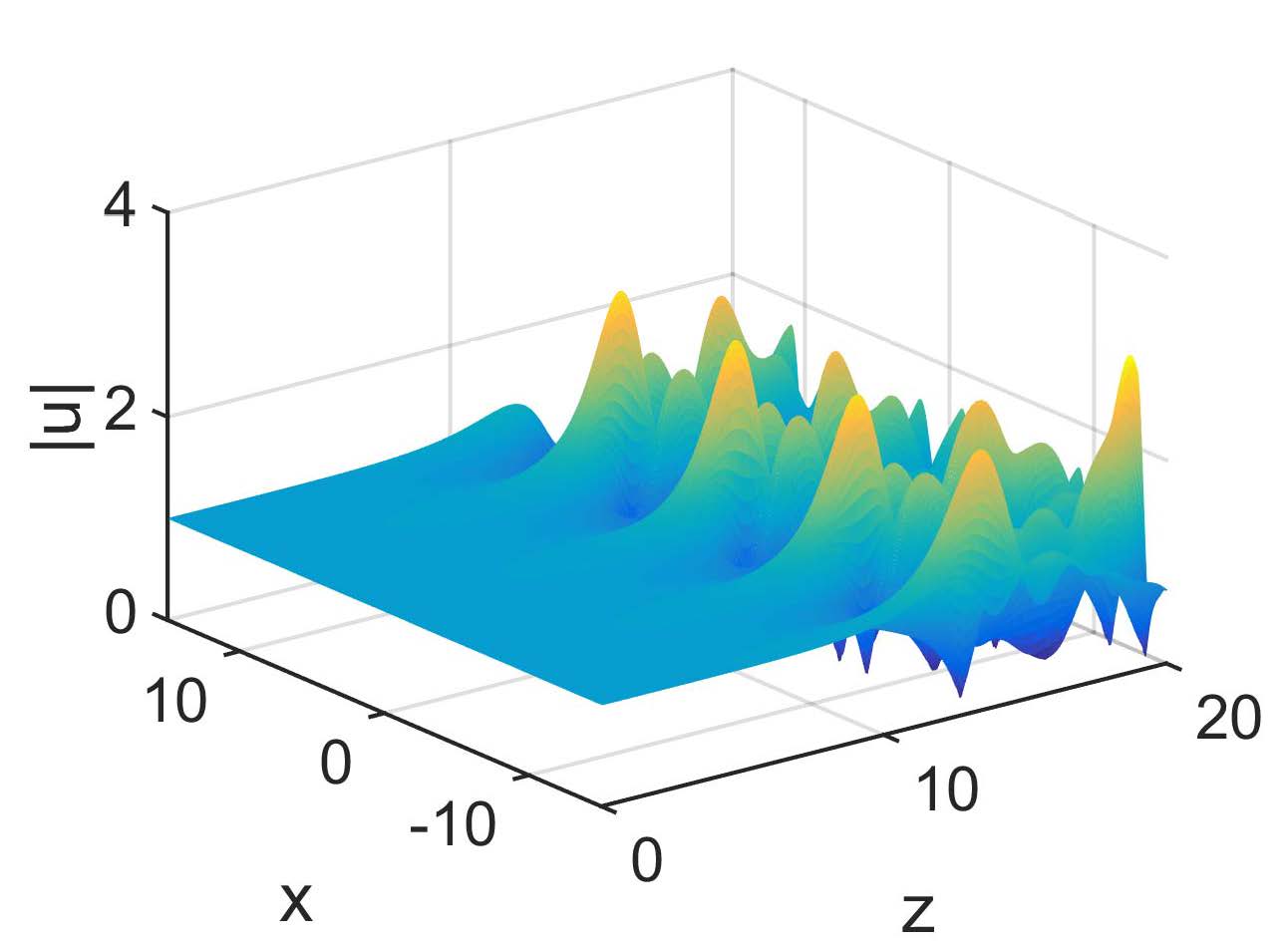}
\includegraphics[height=5.5cm]{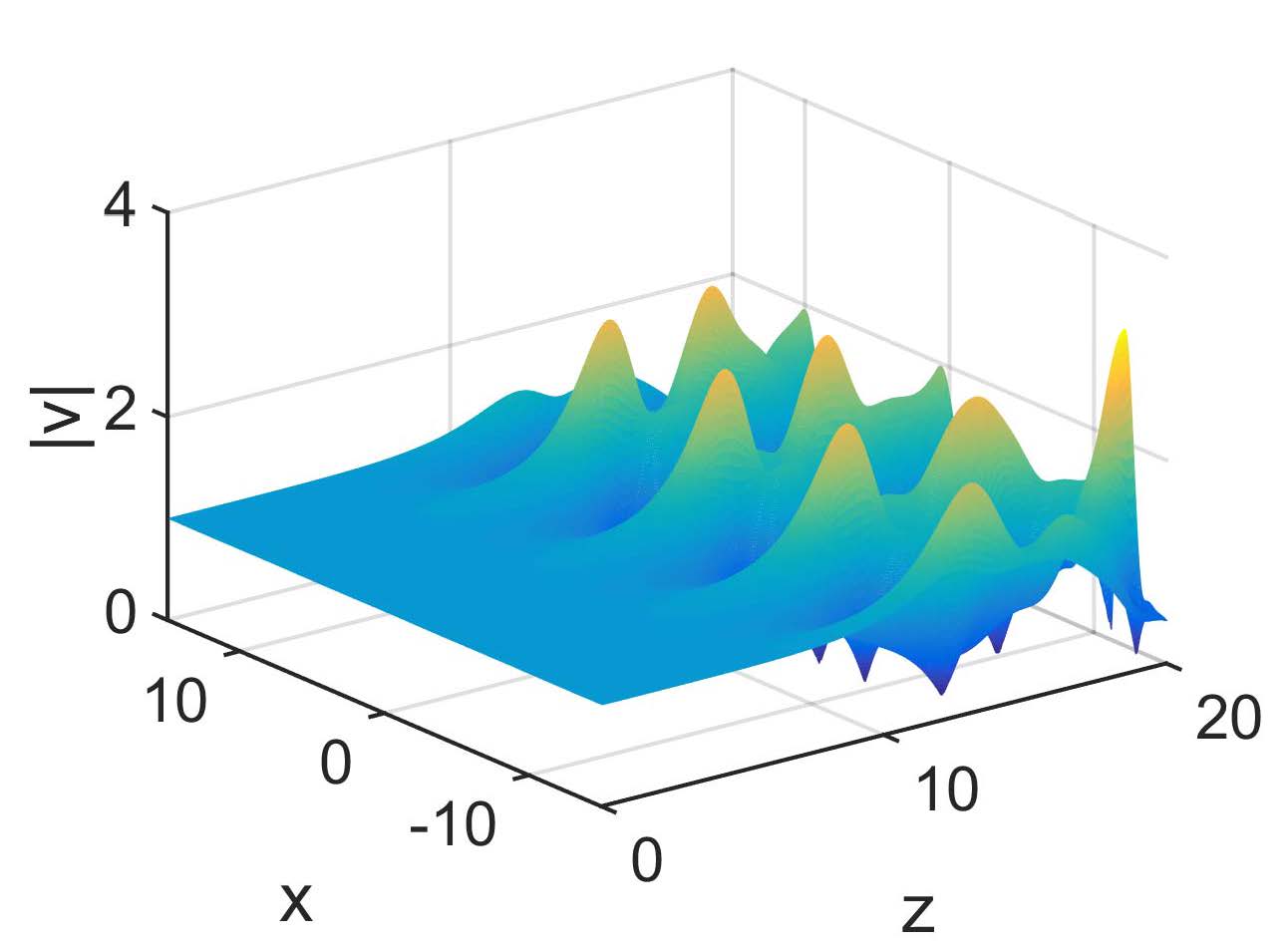}
\caption{(Color Online) Unstable evolution under Eqs. (\ref{nem1})--(\ref{nem3}).
Here $u_1(x)=v_1(x)=0.001 e^{i\pi x/4}$, $d_1=1/2$, $\nu=10$ and all other of Eq. (\ref{eq0})
parameters are equal to unity.}
\label{mi}
\end{figure}

In Fig. \ref{mi} we depict the unstable evolution of a perturbed unit amplitude wave, as Eqs.
(\ref{eq0}). In order to avoid falling into the scalar case ($u=v$) we take $d_1=1/2$, $\nu=10$ and
all other parameters equal to unity. The computations follow the ETDRK4 method of Ref.
\cite{kassam} as appropriate for stiff problems, since stability issues are discussed. For the
plane wave evolution the computational domain $x\in [-4\pi,4\pi]$, $z\in[0,20]$ is used and the
initial conditions are described in Fig. \ref{mi}.

As clearly seen, the small initial perturbation, $u_1, v_1$, results to the
exponential growth of the constant amplitude background, $u_0,v_0$; for short
times and before the nonlinear terms become important. When $\nu$ is increased
the constant background will keep its shape for longer times as also suggested
by Eq. (\ref{omega}).

\begin{figure}[t]
\centering
\includegraphics[height=5.5cm]{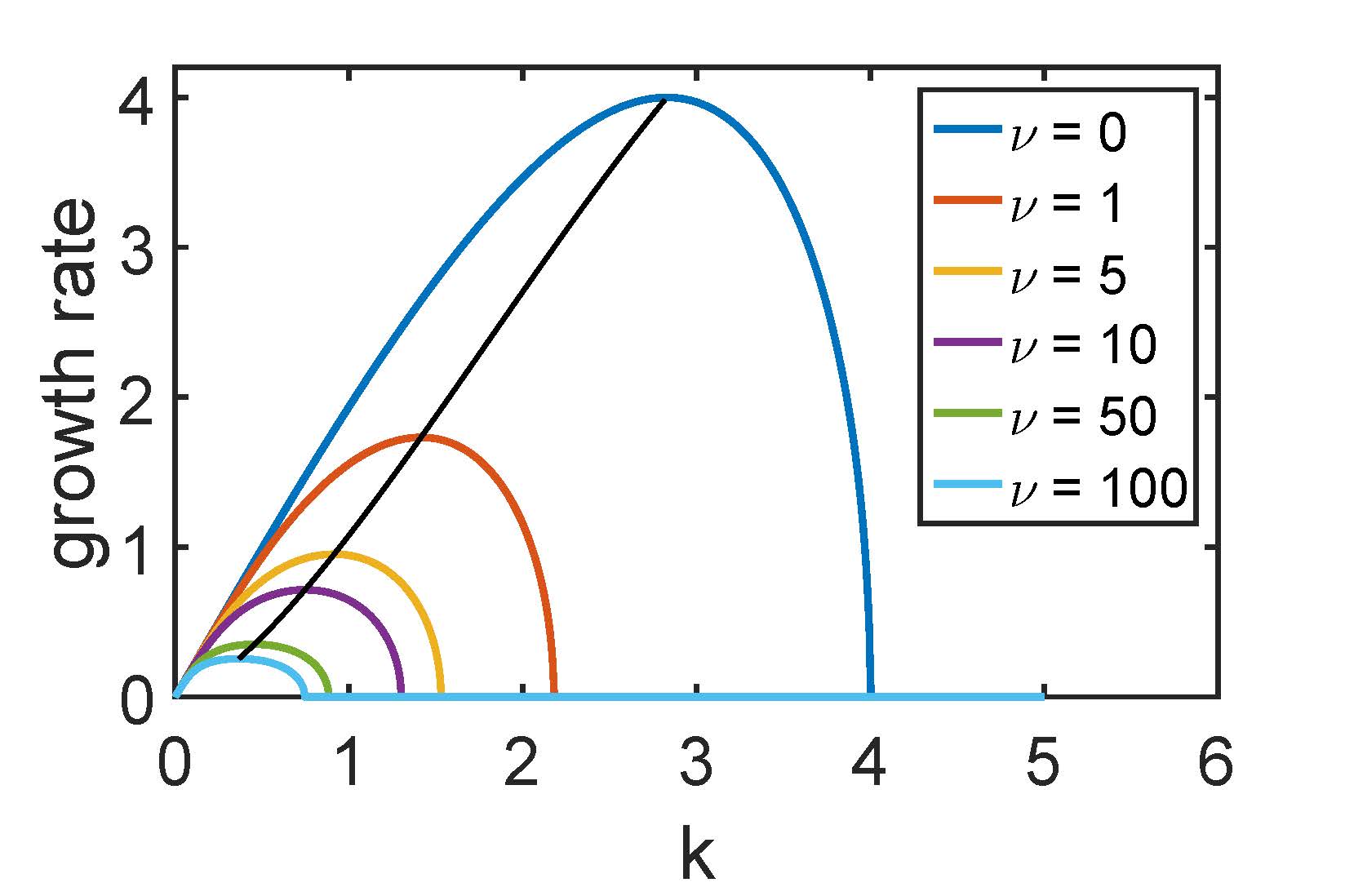}
\caption{(Color Online) The growth rate, Im$\{\omega\}$ of Eq. (\ref{omega}),
of the focusing system for different
values of the nonlocality factor $\nu$. The black solid line corresponds to the maximum growth rate.
All other parameters are kept equal to unity.}
\label{growth.nu}
\end{figure}

We remain on the focusing case where the system is unstable and study the
relative growth rates. As mentioned above, the nonlocal term involving $\nu$
seems to have a stabilizing effect in the sense that growth rates, in the
scalar case, are significantly smaller and MI will need more propagating
distance to occur. This feature is preserved here as well, as seen in Fig.
\ref{growth.nu}. In all figures, the growth rate is defined as $\mathrm{Im}\{
\omega \}$, while $\omega$ is obtained from Eq. (\ref{omega}). We return to
this shortly.

The nonlocal term $\nu$ has a profound effect on the dynamics of plane waves. While the system is
still unstable for large values of $\nu$ the range of wavenumbers that cause instability is
significantly narrower and in addition the maximum growth rate is smaller. That means that for
nematic crystals in particular, where $\nu=O(10^2)$, MI may be suppressed by increasing the value
of the nonlocality or by choosing wavenumbers outside this narrow band that result in unstable
propagation.

\begin{figure}[ht]
\centering
\includegraphics[height=5.5cm]{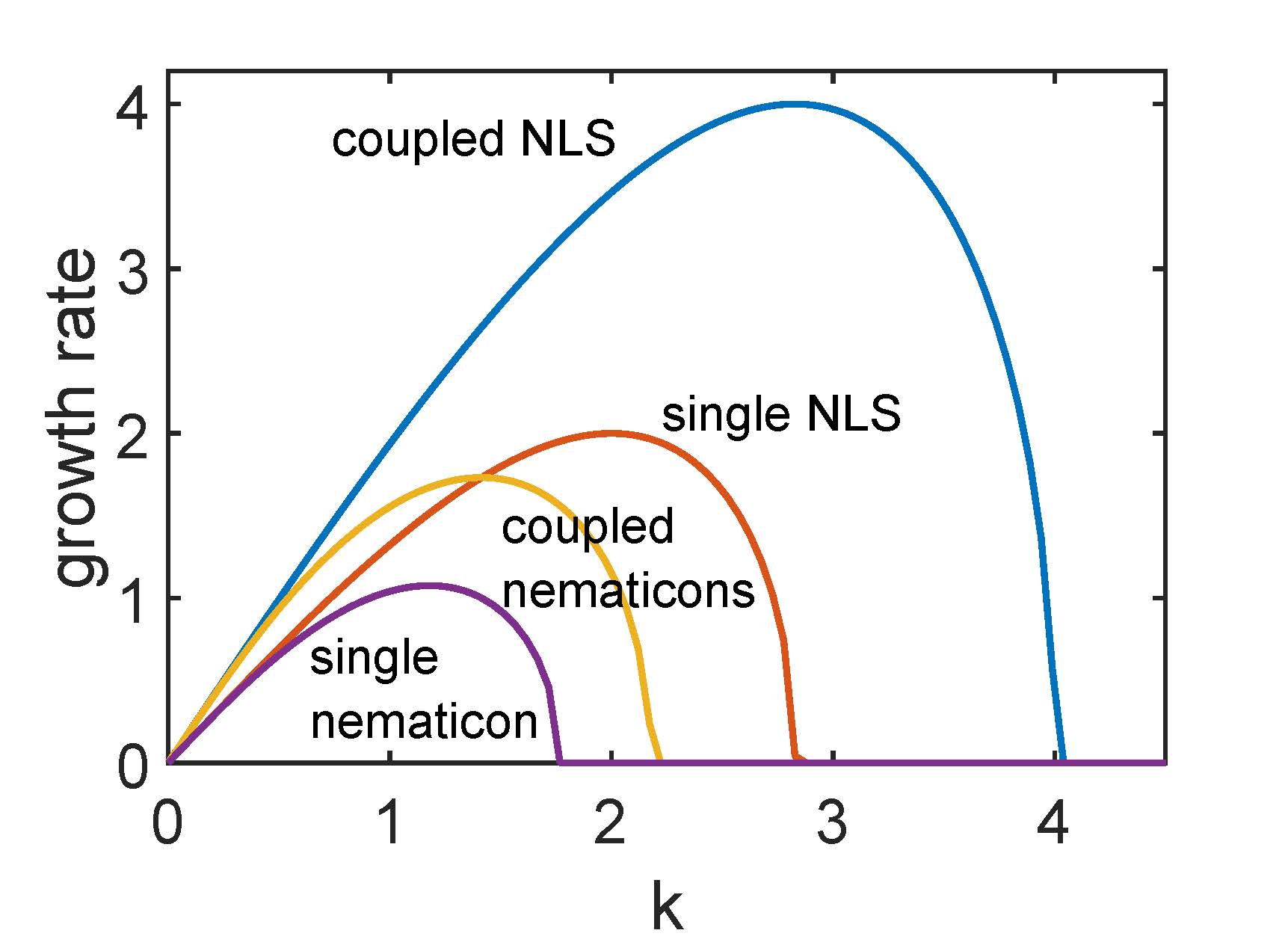}
\caption{(Color Online) Comparison of the growth rates to the relative
NLS equations (Eqs. (\ref{nem1})--(\ref{nem3}) with $\nu=0$). All other parameters are kept equal to unity.}
\label{growth.single}
\end{figure}

However, the coupling provides significantly higher growth rates as seen in Fig.
\ref{growth.single} than the scalar system. As also expected, following this observation, the pure
NLS system ($\nu=0$) has the higher growth rates, so much so that even the single equation
surpasses the coupled nonlocal system -- cf. Fig. \ref{growth.single} (right). Furthermore, and
contrary to the NLS, it has been shown, for the scalar case, that nonlocality of arbitrary shape
can indeed eliminate collapse in all physical dimensions \cite{bang}.

\subsection{The relative NLS system}

The stability criteria found for Eqs. (\ref{nem1})--(\ref{nem3}) are independent of the nonlocal
parameter $\nu$. As such, it is relative to investigate the NLS system that corresponds to the case
$\nu=0$, i.e.
\begin{align}
  i\frac{{\partial u}}{{\partial z}} + \frac{{{d_1}}}{2}\frac{{{\partial ^2}u}}{{\partial {x^2}}}
  + \frac{2}{q}(g_1^2|u{|^2} + {g_1}{g_2}|v{|^2})u &= 0 \label{nls1}\\
  i\frac{{\partial v}}{{\partial z}} + \frac{{{d_2}}}{2}\frac{{{\partial ^2}v}}{{\partial {x^2}}}
  + \frac{2}{q}({g_1}{g_2}|u{|^2} + g_2^2|v{|^2})v &= 0 \label{nls2}
\end{align}
Following the same steps in the analysis above (or simply setting $\nu=0$ everywhere) we obtain the
dispersion relation
\[
32q{\omega ^4} - 8{k^2}[(d_1^2 + d_2^2)q{k^2} - 8({d_1}g_1^2u_0^2 + {d_2}g_2^2v_0^2)]{\omega ^2}
+ {d_1}{d_2}{k^6}[2{d_1}{d_2}q{k^2} - 16({d_2}g_1^2u_0^2 + {d_1}g_2^2v_0^2)] = 0
\]
which can also be shown (albeit in a much simpler manner than before) to have real solutions iff
\[
(\mathrm{i})\; \frac{{g_1^2u_0^2}}{{{d_1}}} + \frac{{g_2^2v_0^2}}{{{d_2}}} < 0
\quad \mathrm{and} \quad
(\mathrm{ii})\; d_1g_1^2u_0^2+d_2g_2^2v_0^2<0
\]
i.e. when $d_1$ and $d_2$ are both negative. This is a major difference between this system and the
coupled NLS equations (with arbitrary coefficients) \cite{kourakis}. The NLS system is richer in
dynamics, as stability may also be obtained even when one of the equations is focusing, or in
contrast, the system may be unstable in the purely defocusing regime. In fact, systems much like
Eqs. (\ref{nls1})--(\ref{nls2}) that are symmetric in the nonlinear coefficients (and $d_1=d_2$)
have been studied extensively in the context of nonlinear optics and birefringent optical fibers,
in particular \cite{agrawal}. Nonetheless, this confirms our findings that indeed the NLS $(\nu=0)$
is included in the stability criteria above.

The case $d_1=d_2$ simplifies the analysis above significantly as the signs of the relative
polynomials are easier to find. The interesting observation is that while the above results still
hold in this case, the numeric value of $d_1=d_2$ becomes irrelevant and only its sign is
important. Also, important is to notice that since these results are independent of the parameter
$\nu$, meaning that the coupled nematicon system and the coupled NLS --with the appropriate
coefficients-- share the same stability conditions. The nonlocality does not affect the stability
criterion, it only contributes to the growth rate, Im$\{\omega\}$, and its effect is to slow down
the occurring instability as shown in Fig. \ref{growth.single}.

\subsection{Critical wavenumbers and nonlocality values}

\begin{figure}[ht]
\centering
\includegraphics[height=5.5cm]{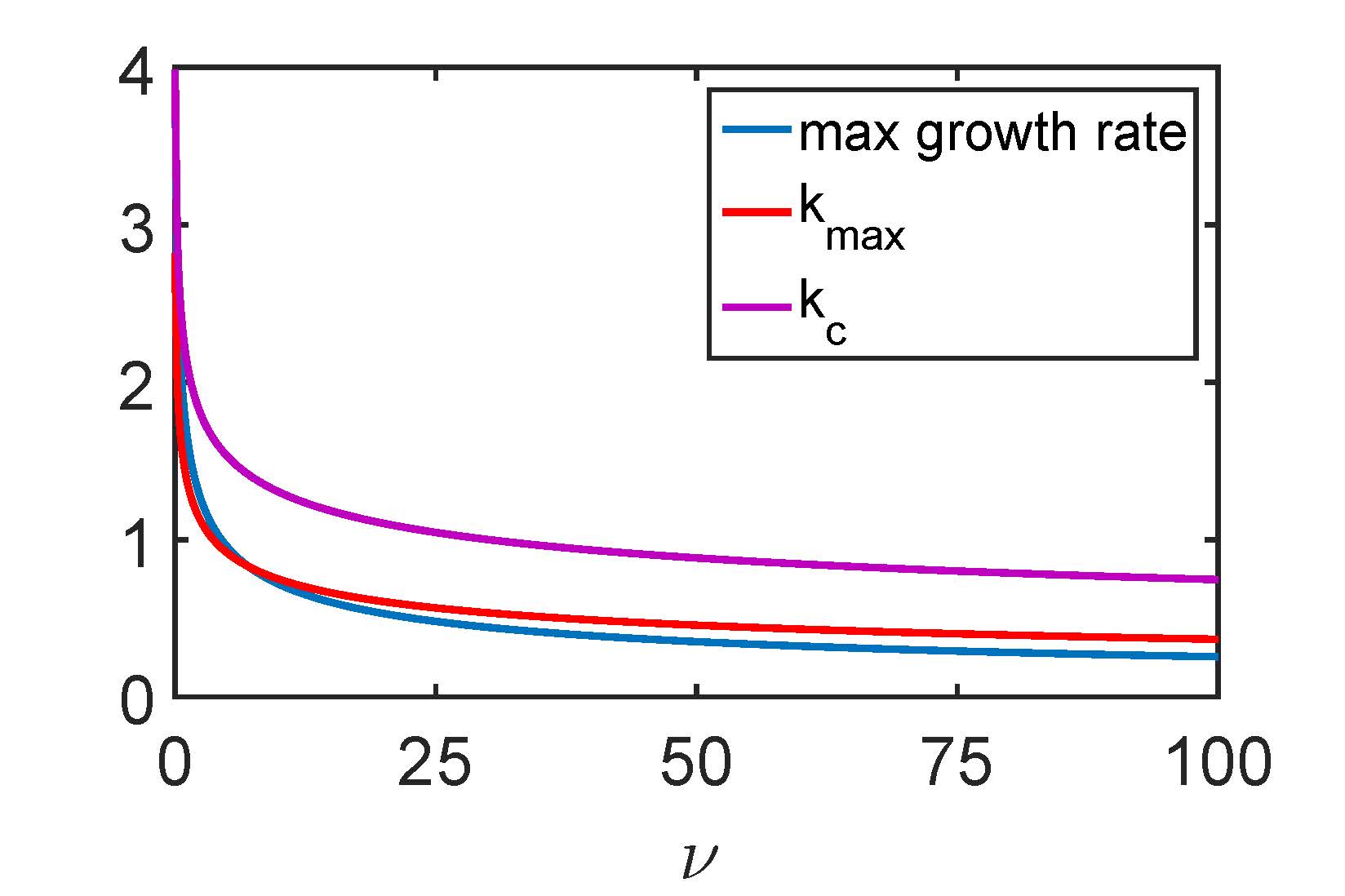}
\caption{(Color Online) Critical wavenumbers and growth rates and their dependence on the
nonlocality.}
\label{nonlocal.k}
\end{figure}

In the MI analysis above, one can identify some critical numbers that play a
key role in the understanding of these results. First and foremost, we identify
the so-called maximum growth rate. This value corresponds to the maxima of Fig.
\ref{growth.nu} and can be found by differentiating Eq. (\ref{omega}), solving
the equation $\omega'(k)=0$ for $k=k_{\mathrm{max}}$ and substituting back to
$\omega_{\mathrm{max}}=\omega(k_{\mathrm{max}})$. The change of
$\mathrm{Im}\{\omega_{\mathrm{max}}\}$ with $\nu$ is shown both in Figs.
\ref{growth.nu} and \ref{nonlocal.k}. However, there is another value that may
be interpreted in two ways. We define a critical wavenumber, $k_c$, which is
essentially the greatest wavenumber for which instabilities can occur. To find
this critical value one needs to solve the inequality below for $k$:
\[
\frac{{ - {p_2}(k) \pm \sqrt {p_2^2(k) - 4{p_1}(k){p_3}(k)} }}{{2{p_1}(k)}} <0.
\]
Then the critical value can be identified as the solution of
\begin{equation}
{d_1}{d_2}\nu {k^4} + 2{d_1}{d_2}q{k^2} - 16({d_2}g_1^2u_0^2 + {d_1}g_2^2v_0^2) = 0.
\label{critical}
\end{equation}
In Fig. \ref{nonlocal.k}, we show the dependence of these critical values with the nonlocality
$\nu$.

In particular, for the values of the figures above we find that
\[
k_c = \sqrt {\frac{{\sqrt {32\nu  + 1} }}{\nu } - \frac{1}{\nu }}
\]
which also demonstrates the way this value is affected by the nonlocality. The relative
$k_{\mathrm{max}}$ (and from that $\omega_{\mathrm{max}}$) value can be obtained from the solution
of the equation ${\nu ^2}{k^6} + 4\nu {k^4} + 4{k^2} - 32 = 0$.

Finally, and following the analysis of Ref. \cite{caplan} one can seek the
critical value of the nonlocality parameter that stabilizes a cw of particular
wavenumber. This is retrieved again from Eq. (\ref{critical}) only now we solve
for $\nu$, i.e.,
\[
\nu  = 2\frac{{ - {d_1}{d_2}q{k^2} + 8({d_2}g_1^2u_0^2 + {d_1}g_2^2v_0^2)}}{{{d_1}{d_2}{k^4}}}
\]
For example, for the values as above, the wave number $k=1$ will always
correspond to a stable cw iff $\nu\ge30$, as also confirmed by Figs.
\ref{growth.nu} and \ref{nonlocal.k}. This critical value is not a general
criterion for stability as it depends on the particular wavenumber. This means
that while the particular critical value of $\nu$ may stabilize a specific
wavenumber another value will render the system unstable, consistently with the
analysis above. Only when the system is full defocusing MI is absent.

\section{Solitons}

The dynamics of two-color nematicons propagation and interactions in the nonlocal limit is usually
studied using variational method based on an appropriate trial function/anzatz whose parameters
(amplitude, width, etc) are chosen so that the Lagrangian of the system is minimized
\cite{skuse3,worthy,skuse2}. However, since these are not exacts solutions they are expected to
shed diffractive radiation much like the solutions of the regular NLS system. We intend to remedy
this by finding exact solutions to Eqs. (\ref{nem1})--(\ref{nem3}) and the conditions associated
with these solutions.

As seen above the coupled system is, in the focusing case, unstable. This means
that any initial condition is subject to instability.  Thus, it is natural to
seek conditions under which soliton solutions may exist that will not undergo
the instability process. To find these solutions (if they exist), we assume
that the stationary solutions of the system (\ref{nem1})--(\ref{nem3}) take the
form
\[
\hspace{-1.5cm} u(z,x)=a_1\,\sech^2(bx)e^{i\mu_1 z},\quad v(z,x)=a_2\,\sech^2(bx)e^{i\mu_2 z}, \quad
\theta(x)=a_3\,\sech^2(bx).
\]
Substituting directly into Eqs. (\ref{nem1})--(\ref{nem3}), we obtain the expressions for the
soliton parameters:
\begin{equation}\label{cond1}
\mu_1=\frac{q d_1}{\nu},\quad \mu_2=\frac{q d_2}{\nu},\quad b=\sqrt{\frac{q}{2\nu}},\quad
a_3=\frac{3q\lambda}{4\nu}.
\end{equation}
The solitons' amplitudes are related through
\begin{equation}
g_1 a_1^2 + g_2 a_2^2=\lambda \frac{9q^2}{8\nu}
\label{amps}
\end{equation}
subject to the condition
\begin{equation}
\frac{d_1}{g_1}=\frac{d_2}{g_2}=\lambda.
\label{cond2}
\end{equation}
Although the freedom of one free parameter (which also relates amplitude and velocity in the NLS
case) is not redeemed, another property is obtained. We are now able to control the amplitude of
one of the components through Eq. (\ref{amps}). However, this is again a fundamental difference
with the NLS system. Indeed, in the NLS equations it is straightforward to obtain a variety of
different cases and soliton types (bright and/or dark). Here, only the focusing case can produce
bright solitons while we where not able to find dark solitons in this manner \cite{djf2}. They may,
however, be obtained, in the small amplitude limit, using the methods of Ref.
\cite{small_amplitude}. One final comment is that this procedure may also be used for more than two
equations relating the relative amplitudes through an equation of the form of Eq. (\ref{amps}).

The role of the nonlocal term $\nu$ is profound here as well. In particular, when $\nu\gg 1$, as is
the case for liquid crystals, it may become (experimentally) more difficult to obtain soliton
solutions and radiation free propagation. Indeed, from Eq. (\ref{amps}) the smaller the right hand
side becomes the more difficult it becomes to obtain amplitudes for soliton propagation. In fact,
when this term vanishes it results in $a_1=a_2=0$, i.e. no soliton solutions exist. Furthermore,
Eq. (\ref{amps}) suggests that
\[
a_1^2 \leq \lambda \frac{9q^2}{8\nu g_1}, \; a_2^2 \leq \lambda \frac{9q^2}{8\nu g_2},
\]
meaning that even with the freedom to choose one of the amplitudes that cannot exceed this maximum
value. With $\nu \gg 1$ it is further implied that solitons can only exist in the small amplitude
limit, thus suggesting that for these systems the analysis of Ref. \cite{small_amplitude} may be
more appropriate.

\begin{figure}[ht]
\centering
\includegraphics[height=5.5cm]{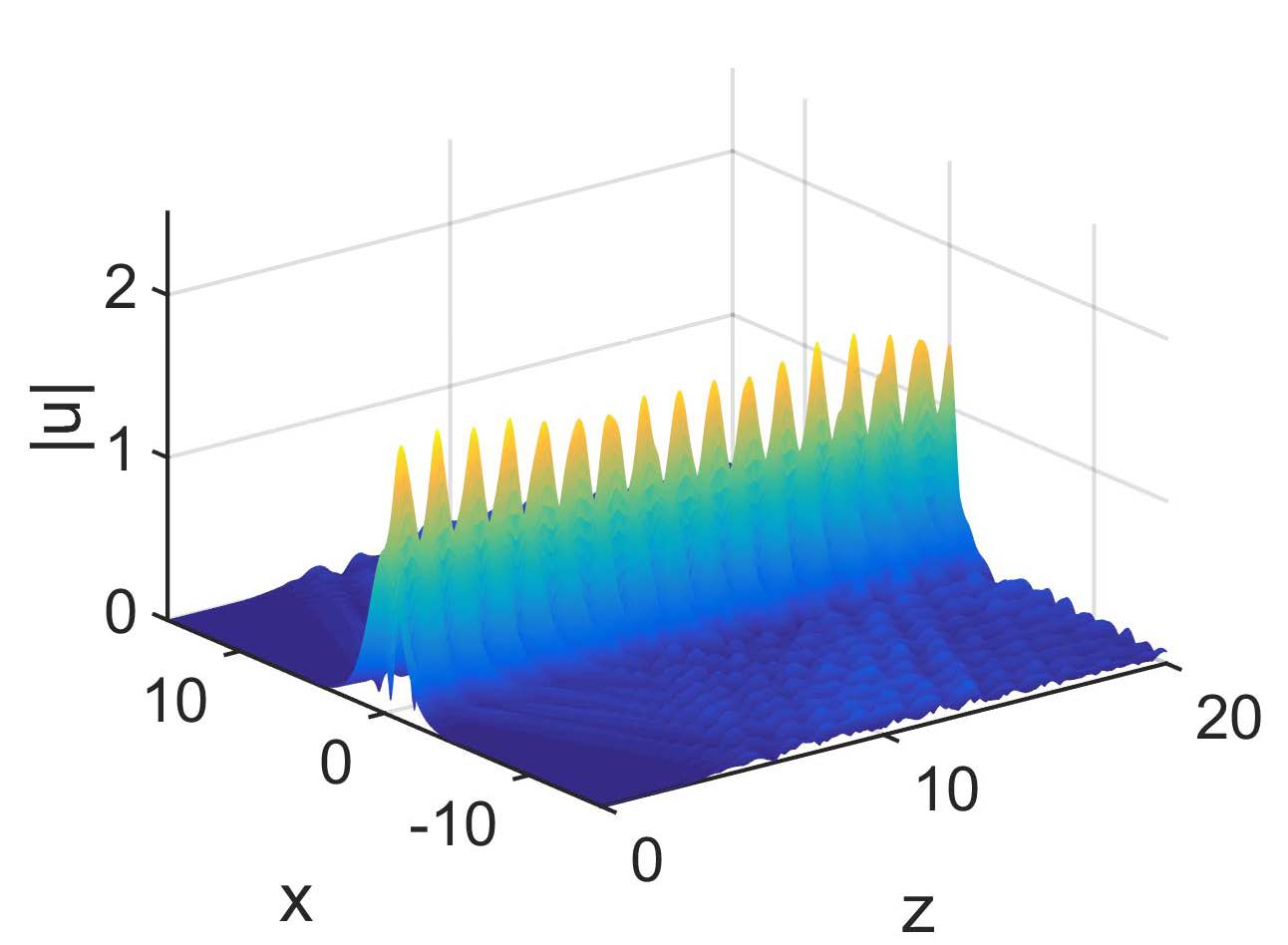}
\includegraphics[height=5cm]{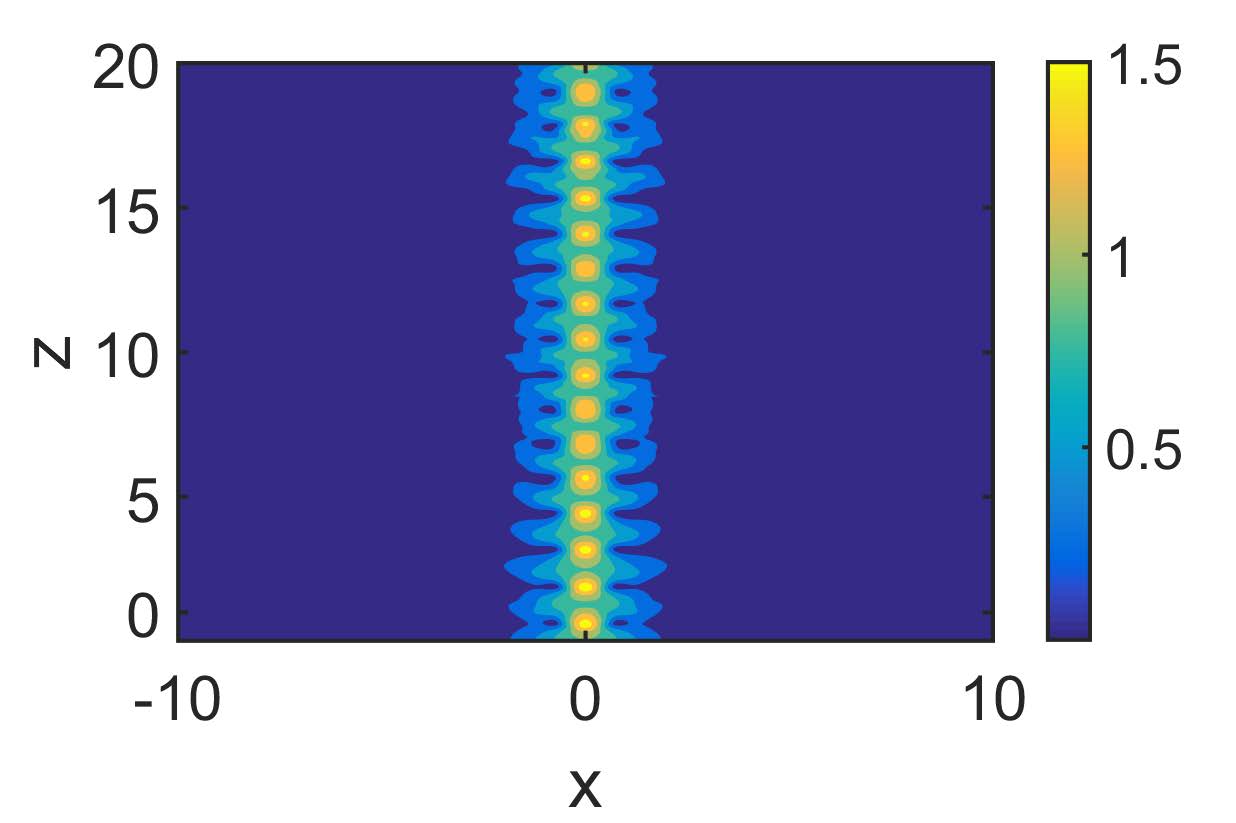}\\
\includegraphics[height=5.5cm]{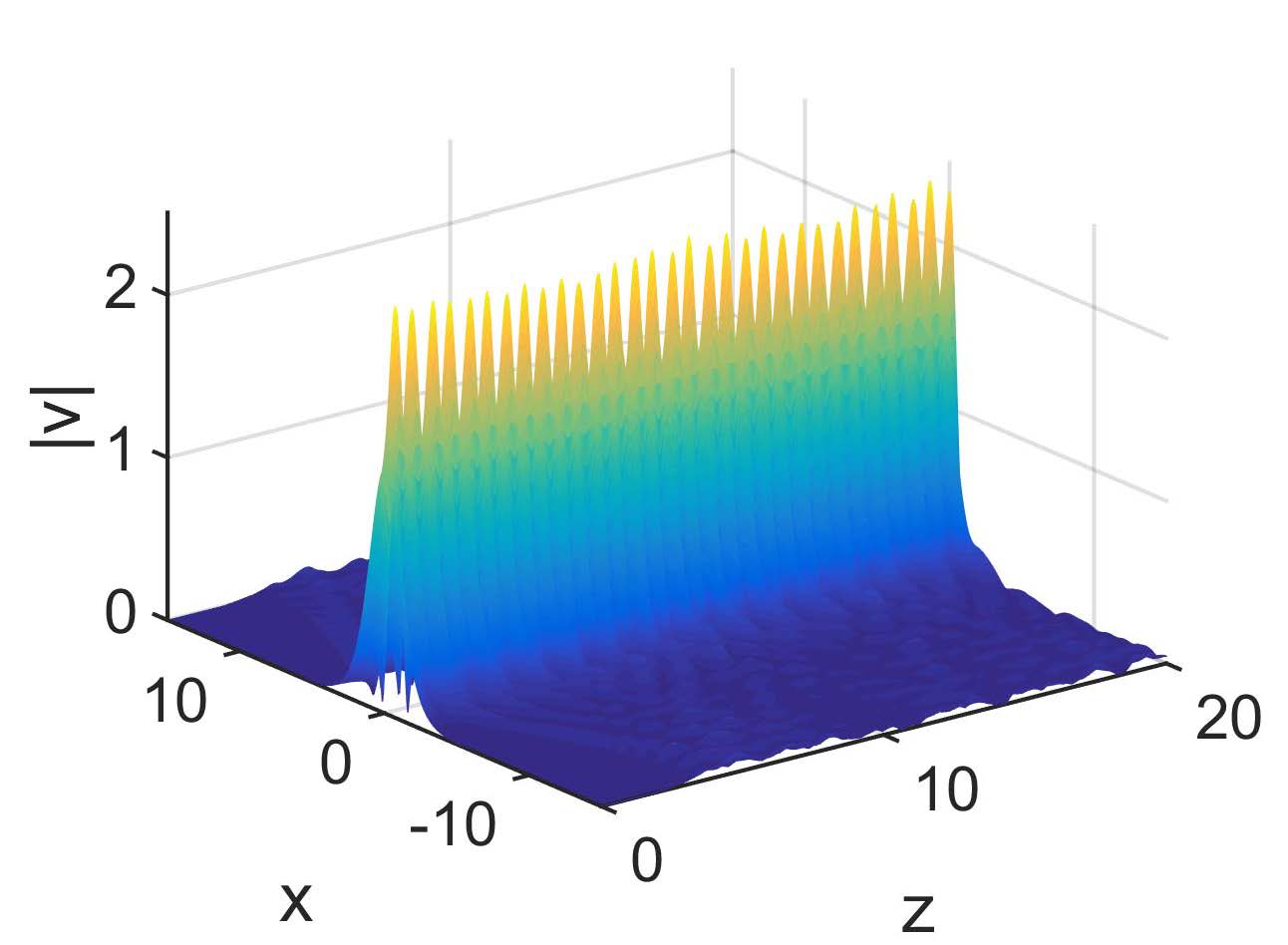}
\includegraphics[height=5cm]{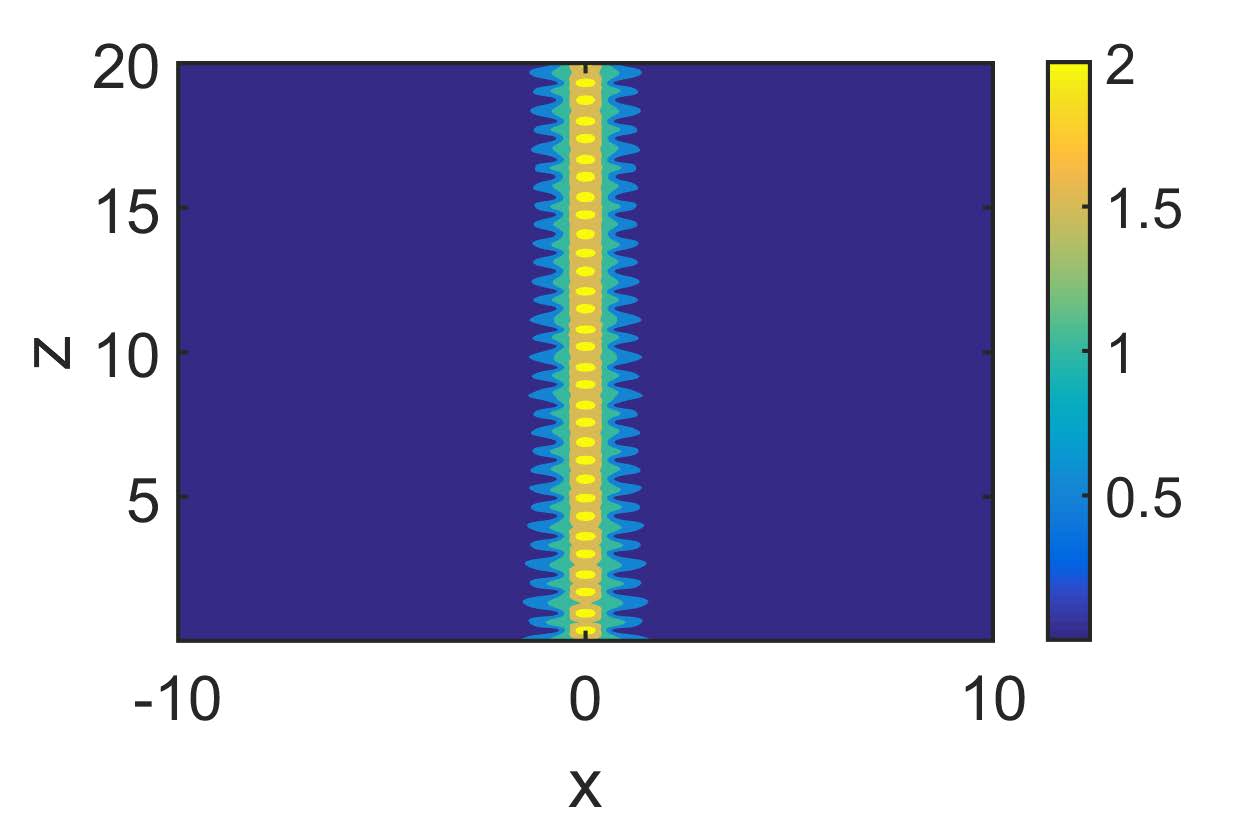}
\caption{(Color Online) A soliton evolution undergoing a breathing behavior. Here we used $d_1=g_1=1$,
$d_2=g_2=2$, $\nu=1$, $q=1$, $a_1=1$ and $a_2=1.5$.}
\label{solitons}
\end{figure}

To illustrate the difference we evolve a pulse which does not obey the amplitude relation Eq.
(\ref{amps}) in Fig. \ref{solitons}.

\begin{figure}[ht]
\centering
\includegraphics[height=5.5cm]{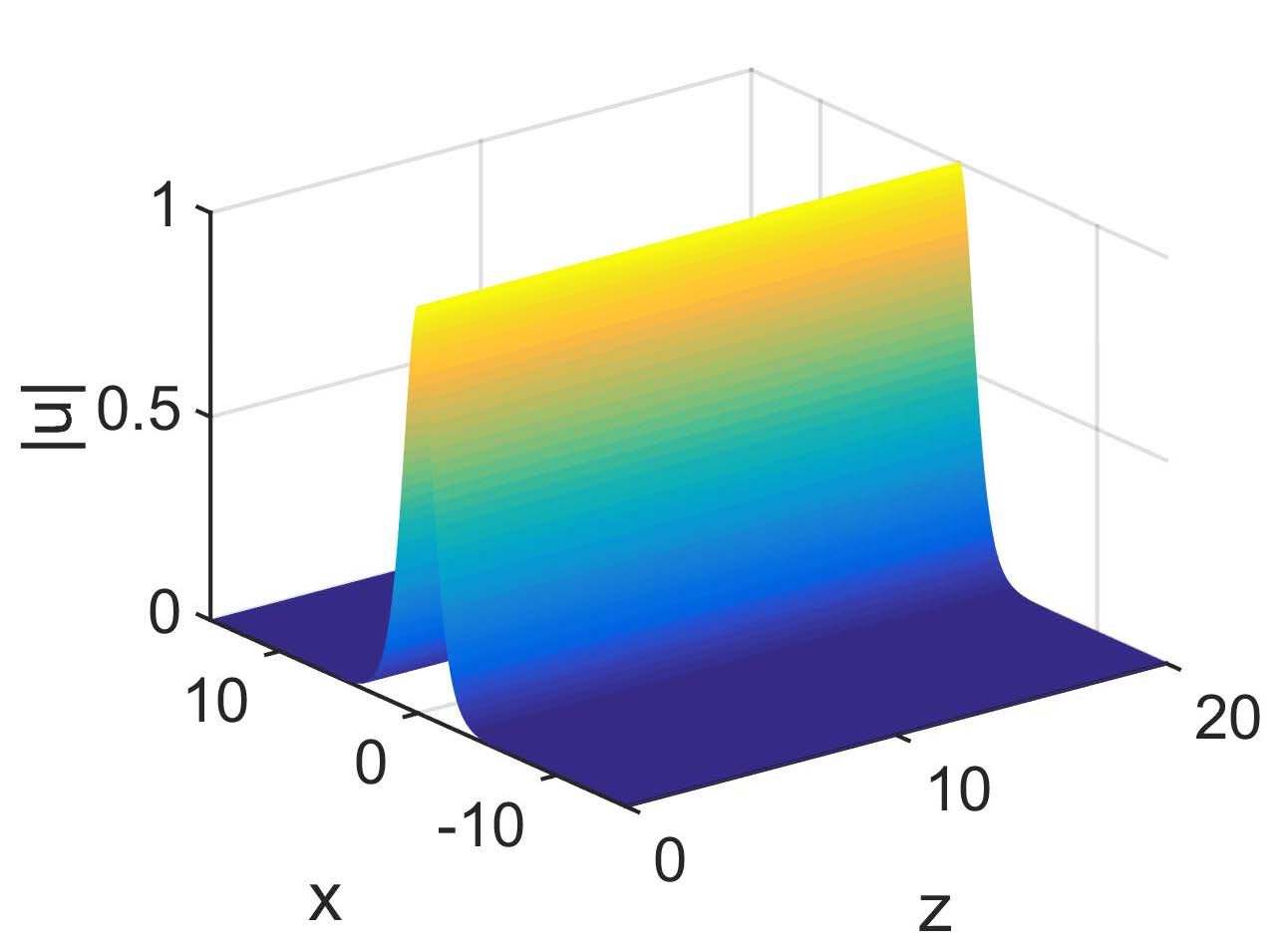}
\includegraphics[height=5cm]{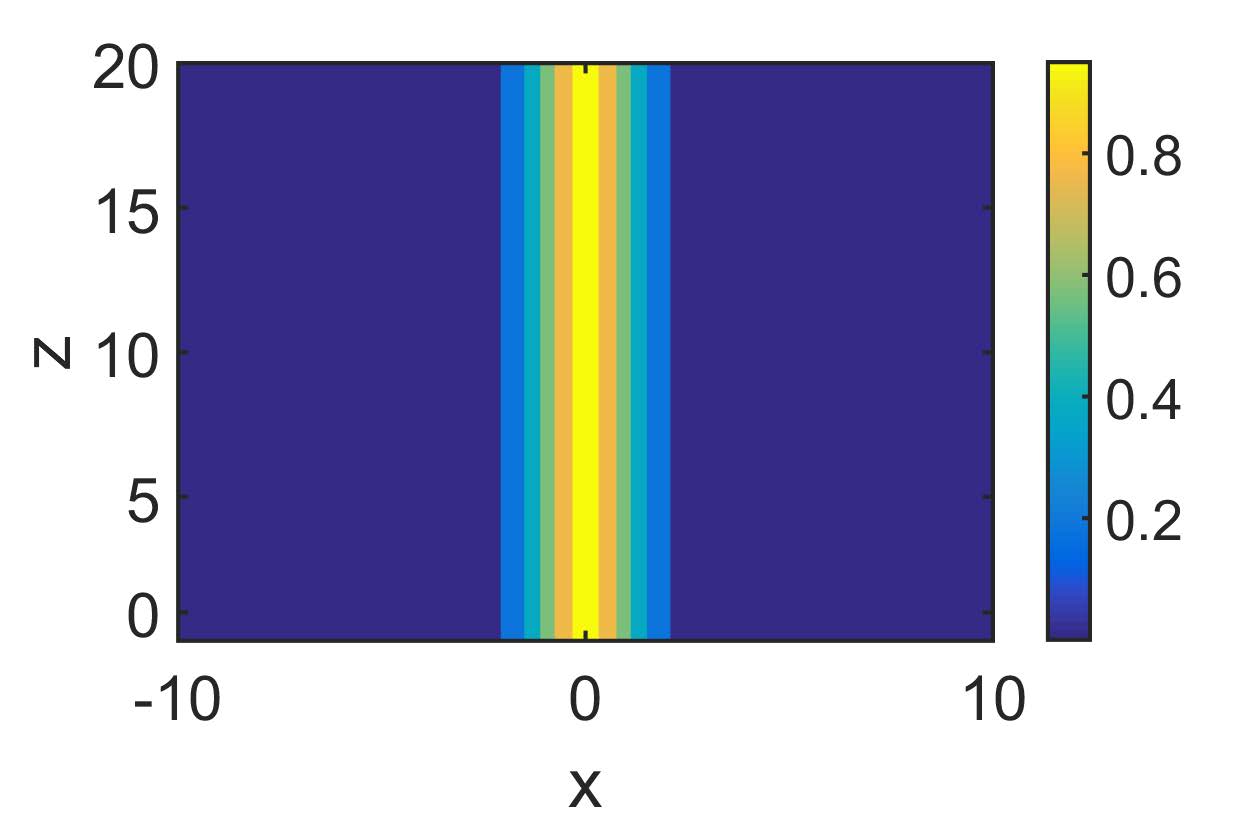}\\
\includegraphics[height=5.5cm]{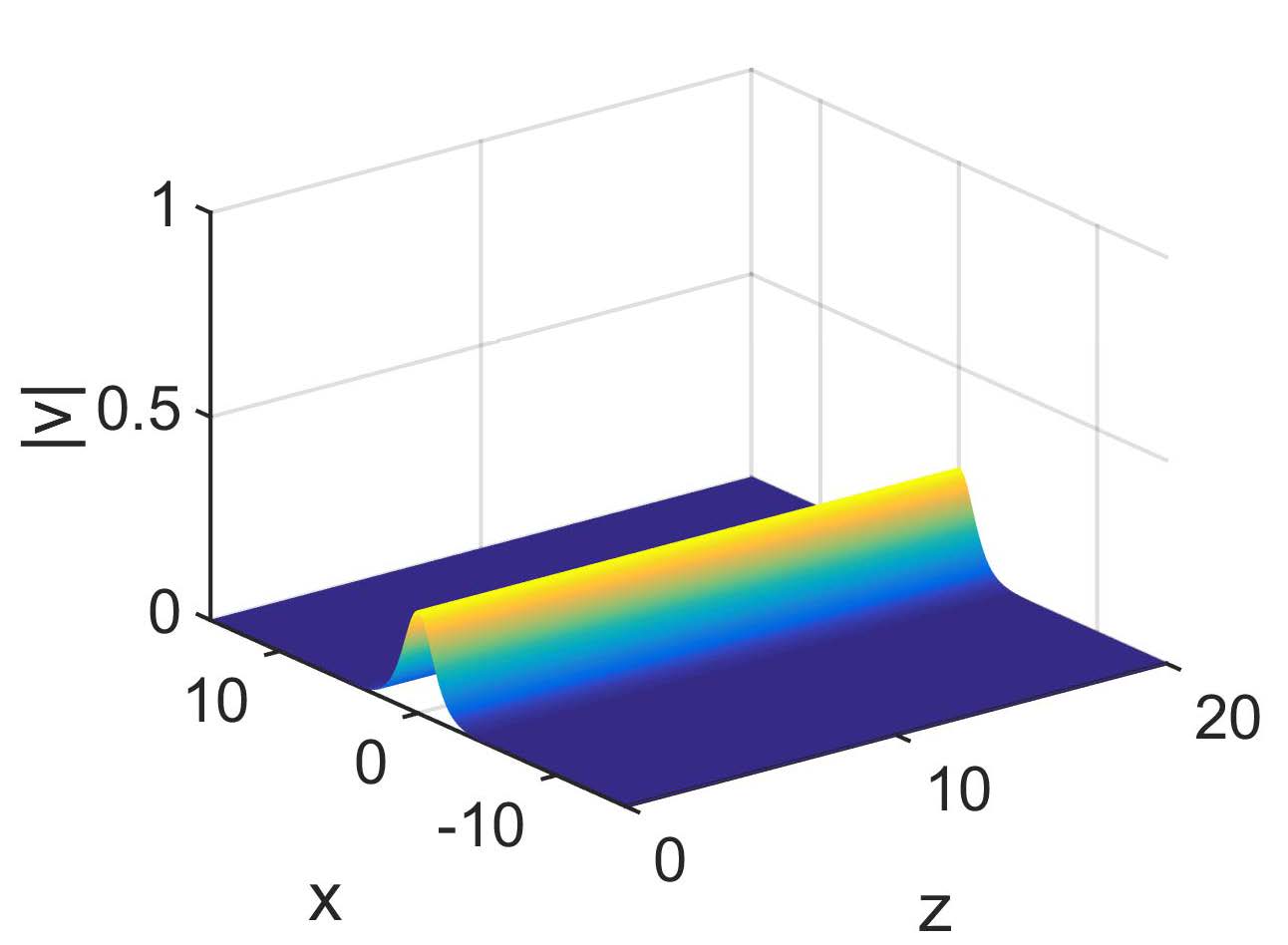}
\includegraphics[height=5cm]{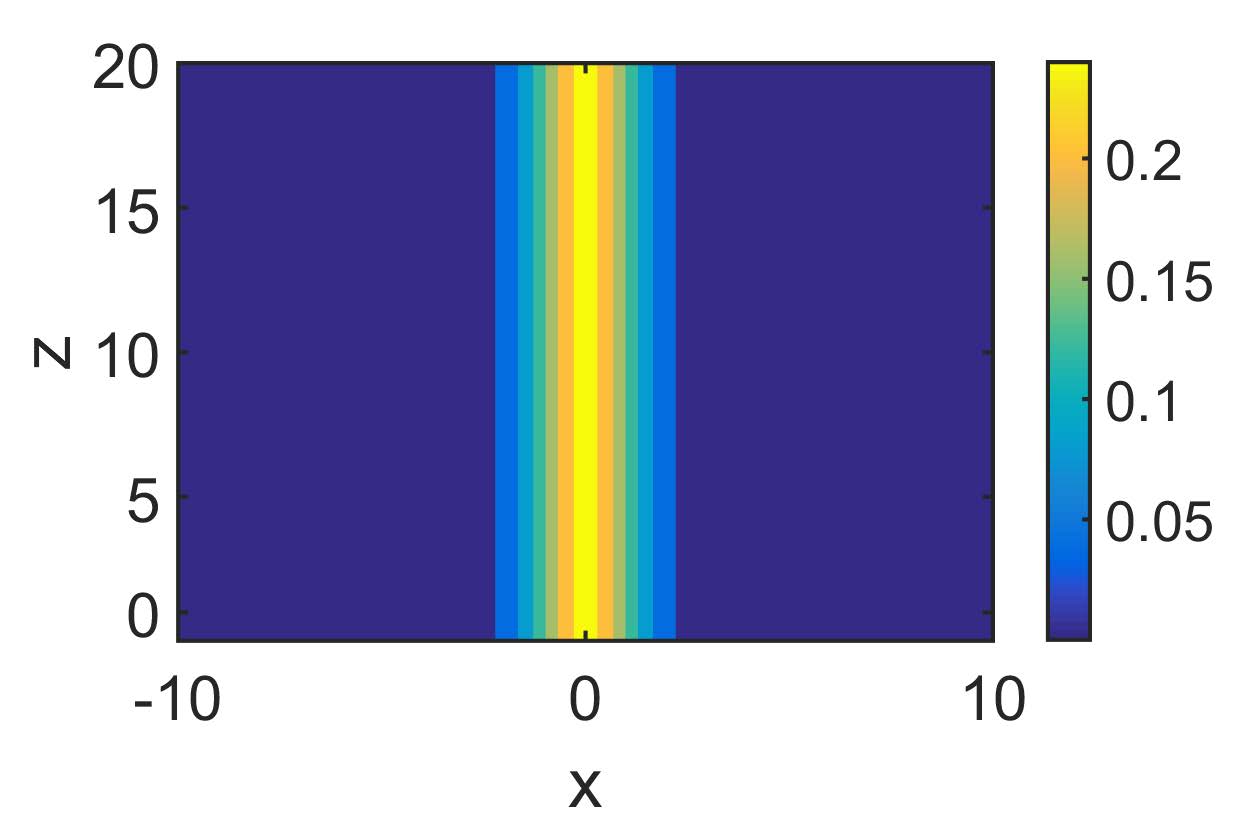}
\caption{(Color Online) A typical soliton evolution. Parameters are same as in
Fig. \ref{solitons} only now $a_2$ is obtained from Eq. (\ref{amps}).}
\label{solitons2}
\end{figure}

Notice here that the initial condition almost immediately starts a breathing behavior and shedding
of radiation. On the other hand if we choose initial conditions that obey the amplitude condition,
Eq. (\ref{amps}), the result is a stable typical solitonic evolution as shown in Fig.
\ref{solitons2}.

\begin{figure}[ht]
\centering
\includegraphics[height=5.4cm]{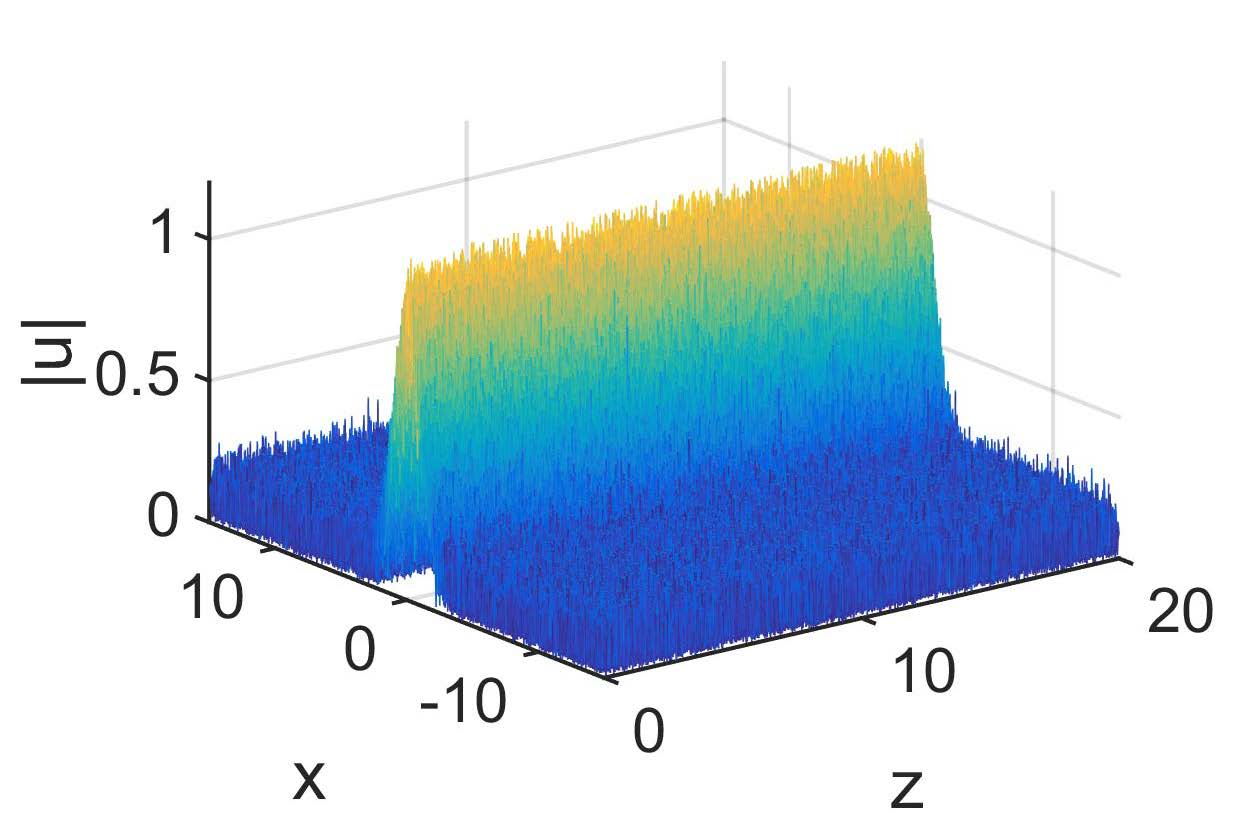}
\includegraphics[height=4.9cm]{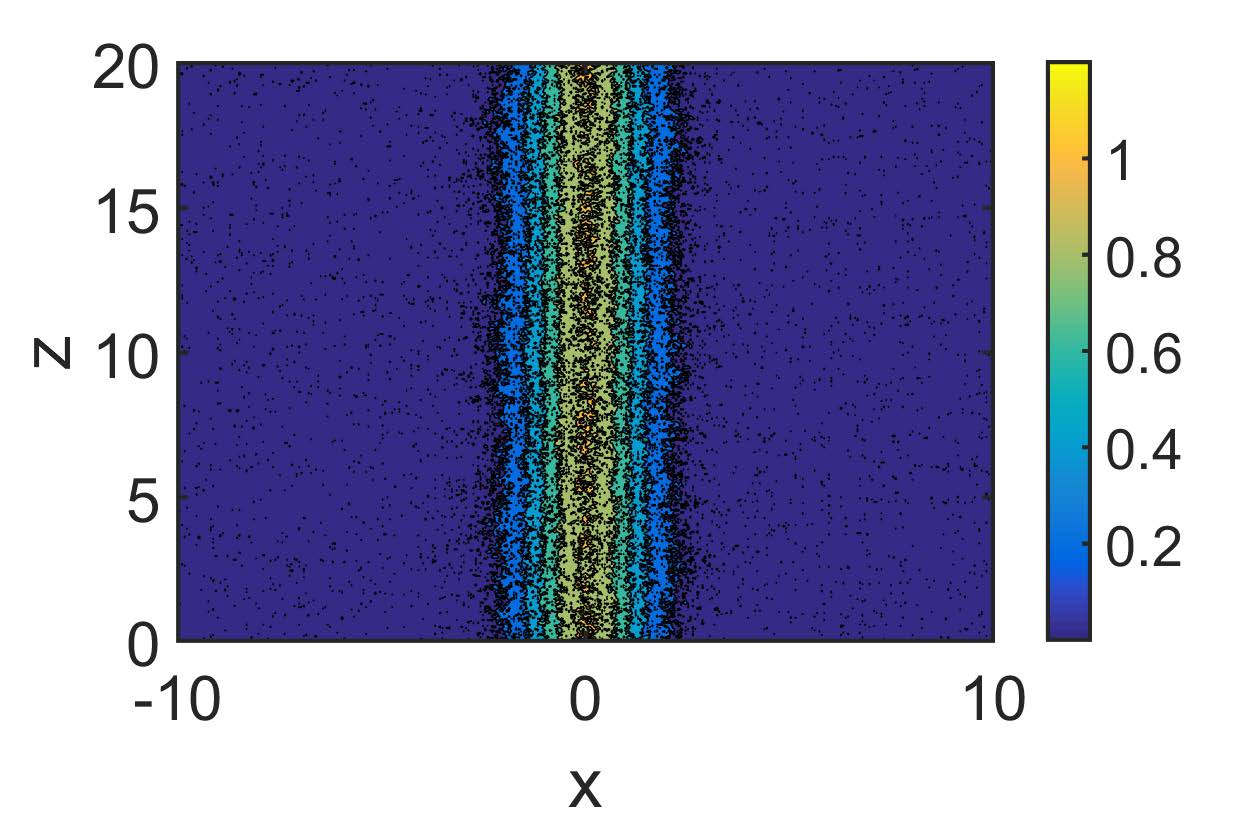}\\
\includegraphics[height=5.4cm]{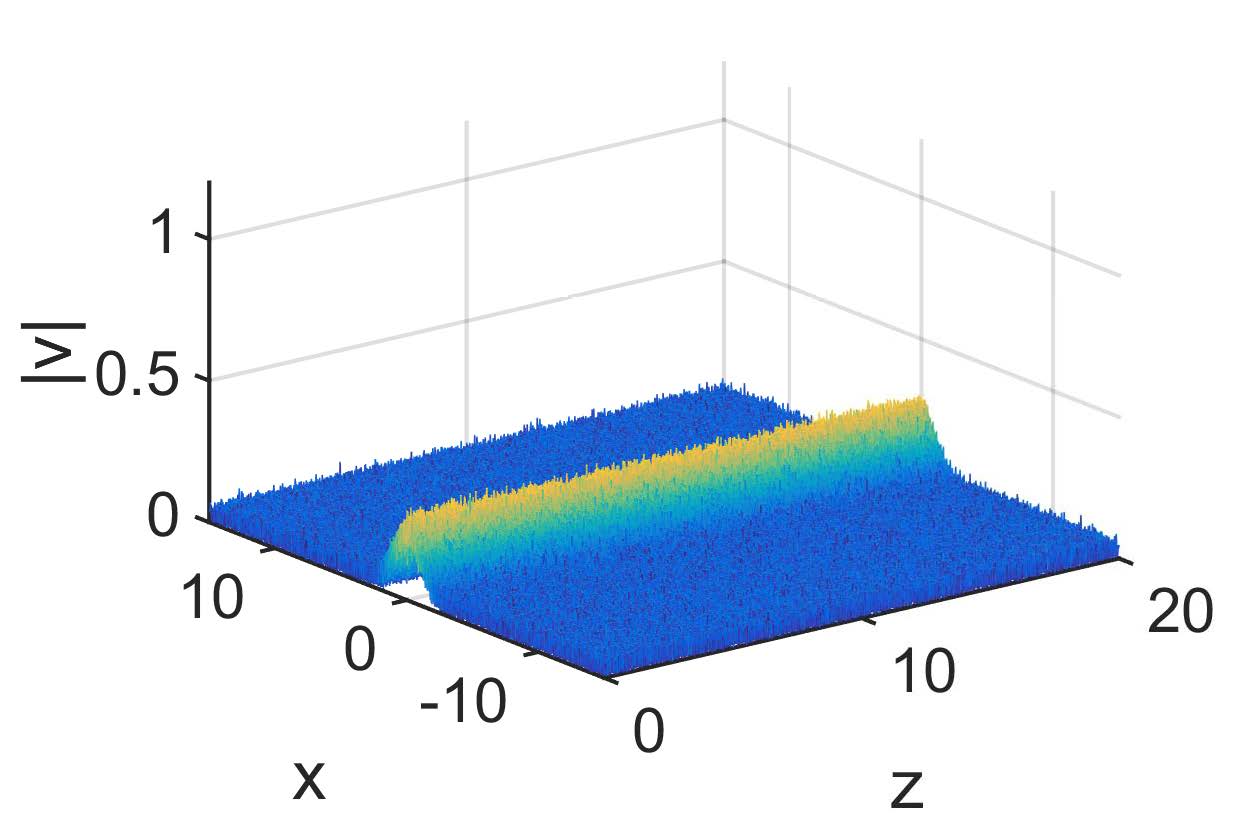}
\includegraphics[height=4.9cm]{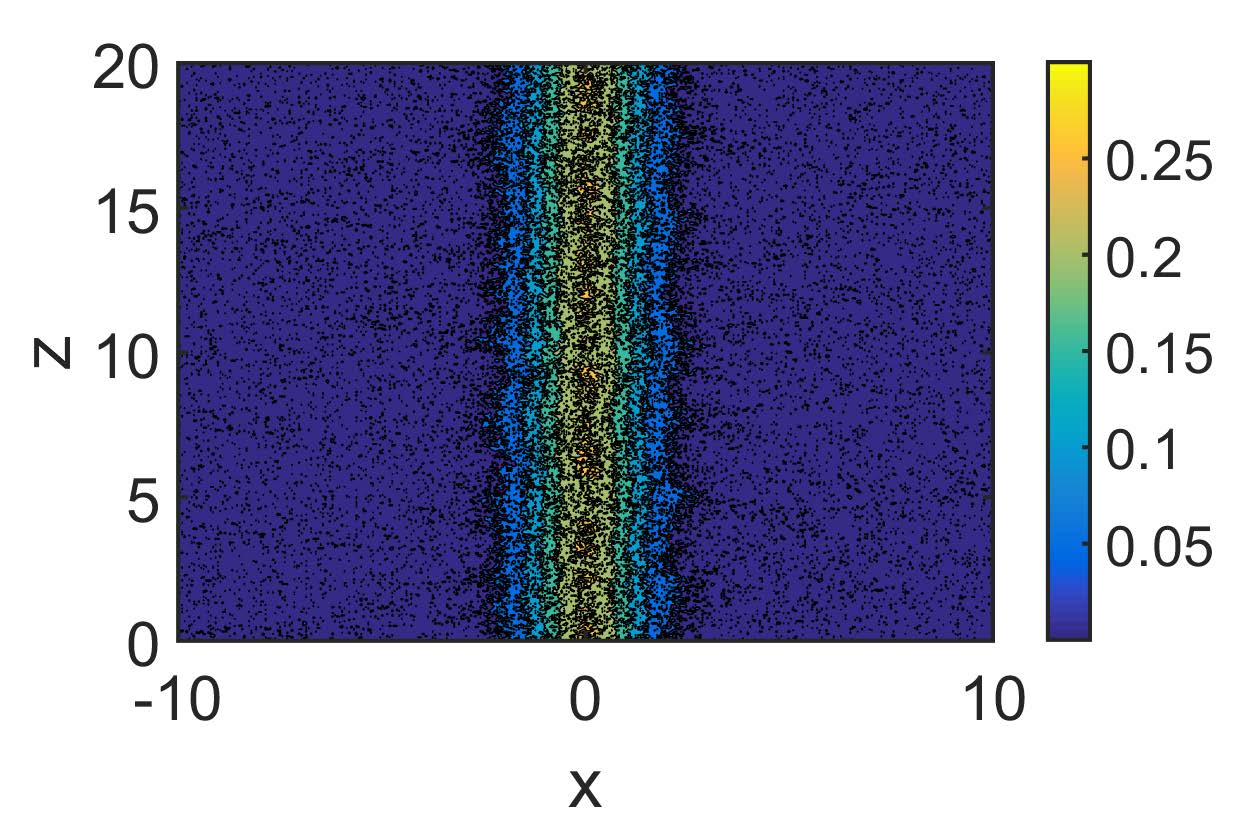}
\caption{(Color Online) The soliton evolution of Fig. \ref{solitons2} with 20\% random noise added,
to numerically test the stability of these solutions.}
\label{solitons.noise}
\end{figure}

To numerically test the robustness of these structures we repeat the above calculation with the
initial conditions of Fig. \ref{solitons2} with 20\% random noise added. The resulting propagation
is depicted in Fig. \ref{solitons.noise}. Notice, that despite the initial noise the pulses
maintain their structural integrity and propagate without radiating. This is strong evidence that
these structure are stable; the complete stability analysis will be provided in a later
communication.

Note, that there is a qualitative difference between the evolutions depicted in Figs.
\ref{solitons} and \ref{solitons.noise}. In the first, where a pulse that does not obey the
amplitude condition, undergoes a breathing behavior but it is not dispersed into radiation because
of the nonlocal nature of the system. In the second, the soliton is exhibiting breathing, but at
the amplitude of the noise making the propagation neutrally stable. This feature can be used to
distinguish pulses in the experimental realization, as noise is an unavoidable part of any
experiment.

\section{Conclusions}

To conclude, we have derived the appropriate conditions for the modulation instability of plane
waves propagating in nonlocal media. It is found that the stability properties follow the ones of
the single system while growth rates are significantly higher. In addition, much like the scalar
case, the nonlocality has a profound effect in the suppression of the effect as it results in
significantly smaller growth rates. Useful information have also been extracted from the study of
the critical values of the analysis: the maximum growth rate and the critical wavenumber which
defines the range of wavenumbers that can induce the instability. In addition, we defined another
critical value based on the nonlocality parameter, which defines the range of values that can
stabilize a cw solution of a particular wavenumber.

In the soliton case, we find that bright solitons may exist strictly on the focusing case and only
when a particular condition holds for the relative amplitudes. Any other pulse will undergo a
breathing behavior but will not disperse into radiation due to the nonlocal nature of the
equations. This condition allows for soliton mutual guiding and may help in their experimental
realization as the correct soliton perturbed with noise will exhibit breathing, but at the
amplitude of the noise making the propagation neutrally stable.

\section*{Acknowledgments}

I wish to thank Stavros Papadakis for his help with the algebra of polynomials. I also thank the
anonymous referees for many helpful suggestions.

\bibliographystyle{elsarticle-num}

\end{document}